\documentclass[12pt]{article}

\usepackage{epsfig,amsmath}

\title{Dirac Observables and the Phase Space of General Relativity}
\author{Hossein Farajollahi \& Hugh Luckock\\ {\small School of
Mathematics and Statistics}\\
{\small University of Sydney}\\
{\small NSW 2006, Australia} }

\date{}

\begin{document}
\maketitle

\def\ext{{\cal E}}  
\def\sympform{{\omega}}
\def\phase{{\cal P}} 
\def\red{{\cal R}}
\def\sub{{\cal N}} 
\def\group{{\cal G}}

\begin{abstract}
In the canonical approach to general relativity it is customary 
to parametrize the phase space by initial data on spacelike 
hypersurfaces. However, if one seeks a theory dealing with 
observations that can be made by a single localized observer, 
it is natural to use a different description of the phase space. 
This results in a different set of Dirac observables from that
appearing in the conventional formulation. It also suggests a
possible solution to the problem of time, which has been one 
of the obstacles to the development of a satisfactory quantum 
theory of gravity. 
\end{abstract}

\newpage

\def\be{\begin{equation}}
\def\ee{\end{equation}}
\def\bea{\begin{eqnarray}}
\def\eea{\end{eqnarray}}
\def\M{{{\cal M}}}
\def\bdy{{\partial\cal M}}
\def\w{\widehat}
\def\n{\widetilde}
\def\real{{\bf R}}
\def\sn{\section}
\def\ssn{\subsection}

\sn{Introduction}
The development of a satisfactory quantum theory of gravity has 
been hindered by both technical and conceptual difficulties.
The latter include the problem of time and the closely related problem 
of identifying suitable observables in general relativity, both of 
of which arise from the general covariance of the classical 
theory and reflect the inherent ambiguity in the identification of 
points in space-time \cite{Isham}. 

In a relativistic theory, an observer can only measure quantities inside 
his causal past. Unless the observer happens to live in a deterministic 
space-time \cite{BudicSachs}, such measurements can never provide him 
with sufficient data to predict the future evolution of any system unless 
he assumes the fields are constrained by a suitable set of boundary 
conditions on some surface outside his causal past. 

In the case of electromagnetism, such assumptions may be justifiable.
Laboratory experiments can easily be shielded from incoming 
electromagnetic radiation, which effectively imposes a boundary 
condition on a surface outside the past light cone. While astrophysical 
and cosmological observations cannot be shielded in quite the same way, 
in fact a natural shield already exists; incoming electromagnetic radiation 
can safely be assumed to have the familiar thermal spectrum of the cosmic 
microwave background, since any radiation produced with the origin of the 
Universe would have subsequently been absorbed by the opaque plasma 
of protons and electrons that filled the Universe for about $10^5$ years. 

The situation is quite different for gravity, for the simple reason that 
it is a universally attractive force that cannot be shielded.
Consequently, no laboratory experiment can be protected from the effects 
of incoming gravitational radiation. Similarly, in the context of 
astrophysical and cosmological observations, it is hard to justify any 
assumptions about the properties of gravitational radiation 
from the nascent Universe. In particular, there is little reason to 
suppose that such radiation will be confined to a narrow range of 
frequencies with a well-understood spectrum like electromagnetic radiation.

Our inability to justify assumptions about incoming radiation suggests 
that, for gravity in particular, we should develop a description of the 
classical that is not reliant on assumed boundary conditions, and which 
deals simply with observations that are accessible to a localized observer.
Of course, there is a price to pay; when we abandon our
assumptions about what goes on outside our causal past, we also 
surrender any possibility of solving the classical equations of 
motion to predict the future. It is tempting to conclude that determinism 
is lost, but in fact this is only half true; a local observer can 
reconstruct a unique solution to the classical equations of motion 
in his causal past using a subset of the observational data accessible 
to him. Thus, the abandonment of 
assumptions about what we cannot see prevents us from predicting the 
future, but still permits deductions about the past.

In this paper we propose an approach to canonical general relativity 
which explicitly incorporates the observer into the theory, and in which 
only those quantities that can be measured by this observer are regarded 
as physical. The remaining degrees of freedom, which are unphysical insofar 
as their values cannot be determined by any experiment, are treated in  
much the same way as unphysical gauge degrees of freedom.

While the analysis presented here is somewhat formal, it suggests a new 
definition of the phase space of general relativity and a novel approach 
to the development of a quantum theory of gravity. One of the main 
advantages of this approach is that (even in the case of pure gravity) 
it leads to a large family of Dirac observables, including one which may 
be interpreted as a time parameter and others associated with the results 
of localized measurements. In the traditional approach to canonical
quantum gravity, some of the most profound conceptual problems are a direct 
consequence of the absence of observables \cite{Isham}. 

\sn{Covariant Phase Space}

Phase space is a fundamental concept in the canonical formulation 
of classical mechanics \cite{HenneauxTeitelboim}. We begin with a summary 
of its geometric properties, and then discuss how the phase space is actually 
defined. 

For a theory with no gauge symmetries, the phase space is generally 
a manifold $\phase$ equipped with a symplectic form; i.e. a 2-form 
$\omega$ that is closed ($d\sympform=0$) and nondegenerate. 
The symplectic form is useful because it establishes a bijective linear 
correspondence between vector fields and 1-forms on $\phase$; 
given any vector field $X$ on $\phase$, the associated 1-form 
is given by $i_X\sympform$ (the contraction of $\sympform$ with $X$). 

The symplectic form $\sympform$ also associates with each $C^1$ function 
$f:\phase\mapsto\real$ a unique vector field $X_f$ such that

\begin{equation}
i_{X_f} \sympform = - df. \label{eq:VecDef}
\end{equation}

In particular, if $h:\phase\to\real$ is the Hamiltonian function,
then $X_h$ generates the flow associated with time translation. 
The symplectic form is invariant under the action of this flow, 
since its Lie derivative with respect to $X_h$ vanishes: 
\begin{equation} 
{\cal L}_{X_h} \sympform \equiv i_{X_h}\, d \sympform + d\, i_{X_h}\sympform = 
0 - ddh = 0.
\end{equation}
(Here we used the closure of $\sympform$, as well as the standard 
identities $dd=0$ and ${\cal L}_X=i_X d +di_X$ 
\cite{GockSchuck}.) 
Making further use of the 
symplectic form $\sympform$, we also define the Poisson bracket of 
two smooth functions $f,g$ on $\phase$ as 
\begin{equation} 
\{ f,g\} \equiv -i_{X_f} dg = i_{X_f} i_{X_g} \sympform = 
-i_{X_g} i_{X_f}\sympform = i_{X_g} df.   
\end{equation} 
In a gauge theory the phase space $\phase$ is {\it not} equipped with 
a symplectic form, and hence there is no Poisson bracket on $\phase$. 
Instead,  $\phase$ is equipped with a {\it presymplectic} form; 
i.e. a 2-form $\pi$ that is closed but degenerate, in the sense that 
$\phase$ admits a non-zero vector field $V$ with $i_V\pi=0$. 
In general, $\pi$ can be regarded as the restriction to $\phase$ of 
a symplectic 2-form $\sympform$ defined on some {\it extended} 
phase space $\ext$ in which $\phase$ is embedded. 

The phase space $\phase$ may be described as the surface in $\ext$
on which a set of constraint functions $\Phi_a:\ext\mapsto\real$ 
vanish. For the sake of brevity, 
we will assume here that all these constraints are independent and 
first-class, which means that they form a closed algebra under the 
action of the Poisson brackets defined on $\ext$ by $\sympform$; i.e.  
\begin{equation}
\{ \Phi_a, \Phi_b\} = f_{ab}{}^c \Phi_c  \label{eq:PoissAlg}
\end{equation}
for a suitable choice of functions $f_{ab}{}^c=-f_{ba}{}^c$ on $\ext$. 
(In general there 
may also be second-class constraints that cannot be included in 
such an algebra. These indicate the presence of redundant degrees 
of freedom, which - at least in principle - can be eliminated 
from the extended phase space $\ext$. Here we will assume this 
has been done.) 

Associated with each first-class constraint function $\Phi_a$ is 
a vector field $V_a$ for which $i_{V_a}\sympform=d\Phi_a$ on $\ext$. 
These vector fields  are tangential to the constraint surface $\phase$ 
since
\begin{equation} 
i_{V_a} d\Phi_b = - i_{V_a} i_{V_b} \sympform =-  \{ \Phi_a,\Phi_b\} =
- f_{ab}{}^c \Phi_c \ \ = 0\ \hbox{on $\phase$}.
\end{equation} 
and correspond to the null directions of the degenerate form $\pi$ on 
$\phase$; 
\begin{equation} 
\left.i_{V_a} \pi\right|_\phase = \left. i_{V_a} \sympform\right|_\phase = 
-\left. d\Phi_a\right|_\phase = 0 \ \ \hbox{since $\Phi_a=0$ on $\phase$.} 
\end{equation}
The vector fields $V_a$ can easily be shown to form a closed algebra 
\begin{equation}  
[V_a,V_b] = - f_{ab}{}^c V_c 
\end{equation} 
and are in fact the generators of the gauge group $\group$. The 
corresponding directions in $\phase$ are thus associated with purely 
gauge degrees of freedom. 
 
These unphysical degrees of freedom can be eliminated by 
identifying points in $\phase$ that can be mapped into each 
other by gauge transformations. One is left with the quotient 
space $\red=\phase/\group$, whose elements are the orbits in 
$\phase$ of the gauge group $\group$. This space inherits a 
non-degenerate symplectic form from $\phase$, and is sometimes 
referred to as the {\it reduced} phase space of the theory 
\cite{HenneauxTeitelboim}. 
It may be thought of as the phase space of just the 
physical degrees of freedom.   

The above discussion provides a useful geometric insight into the 
canonical theory, but it leaves an important question unanswered: 
given the dynamical laws governing a physical system, how does 
one determine what the phase space is? While the answer is often 
straightforward, this is not always the case and so it is useful 
to give a definition that can be applied in any situation. 

In fact the phase space $\phase$ for a classical system can be 
defined in a fully covariant manner as the space of solutions 
to the equations of motion \cite{Witten,Crnkovic,RovelliLandi}.
A single point in phase space is thus identified with an entire solution 
of the equations of motion, rather than with the state of the system at a 
particular instant\footnote{The perspective presented here differs from 
the conventional one in much the same way that the Heisenberg picture 
of quantum mechanics differs from the Schr\"odinger picture.}. The 
evolution of the system is therefore not represented by a Hamiltonian 
flow in phase space, as it is in the standard approach. Instead, 
the Hamiltonian generates transformations which map a given solution to 
a {\it distinct} solution, related to the first by a time translation. 

As with any manifold, there are many possible coordinate systems that 
can be used to label points in $\phase$; i.e. to identify particular 
solutions of the equations of motion.
One natural approach is to label a particular solution by a set of 
initial data from which it can be uniquely determined. 
For example, in the familiar case of a particle moving in 
the configuration space $\real^N$, 
each point in phase space is represented by a classical trajectory;  
i.e. a $C^1$ mapping $\gamma:\real\to\real^N$ satisfying the 
Euler-Lagrange equations. (If symmetries are present there will be 
a family of physically indistinguishable trajectories representing 
the same point in phase space.) 

A natural way to identify the trajectory $\gamma$ is by specifying an 
appropriate set of initial data. For example we could identify 
$\gamma$ by specifying the instantaneous values of the $N$ 
coordinates and their derivatives at the time $t=0$; 
\begin{equation} 
(q^1,\ldots q^N)\equiv \gamma(0), \ \ \ \ \ 
(\dot q^1,\ldots, \dot q^N) \equiv \left. \frac{d\gamma}{dt} \right|_{t=0}.  
\end{equation}
In a typical theory, these initial data would be sufficient to 
uniquely identify the entire trajectory $\gamma$. Another approach 
is to eliminate the velocities $\dot q^i$ in favour of an equal 
number of functions $p_i(q,\dot q)$ called momenta, defined so that 
\begin{equation}
dp_i\wedge dq^i=\sympform.
\end{equation} 
The resulting description appears completely standard, except that 
the 2N-tuple $(q^1,\ldots q^N,p_1,\ldots p_N)$ labels the entire 
classical trajectory $\gamma$ rather than just a point on the 
trajectory. As a consequence, the evolution of the system is not 
described by motion in phase space. 

When the theory admits gauge symmetries, the equations of motion will 
admit 
a number of distinct solutions that are physically indistinguishable. 
Each point in the reduced phase space $\red$ is then taken to represent a 
single equivalence class of such solutions. For example, suppose that 
two classical trajectories $\gamma$ and $\bar\gamma$ can be mapped 
into each other by a gauge transformation, and hence are physically 
indistinguishable. Both $\gamma$ and $\bar\gamma$ are represented by the 
same point in phase space, which is defined as the equivalence class 
$[\gamma]$ of all trajectories that are physically indistinguishable 
from $\gamma$. 

A special kind of symmetry is that associated with reparametrization.
In a theory of the kind described above, a trajectory 
$\bar\gamma:\real\to\real^N$ is said to be related to $\gamma$ by 
a reparametrization if there exists an increasing $C^\infty$ bijection 
$\tau:\real\mapsto\real$ such that 
\begin{equation} 
\bar\gamma(t) =\gamma(\tau(t)) \ \ \ \forall t\in\real.  
\end{equation}
If such trajectories are physically indistinguishable, the theory is 
said to possess reparametrization symmetry. In such a theory, a point 
in phase space is an equivalence class $[\gamma]$ of classical 
trajectories related to each other by reparametrization, and is 
represented by a directed curve in $\real^N$ without any preferred 
parametrization. 

The absence of a natural parametrization introduces additional ambiguity 
into any attempts to label this curve using initial data. In order to 
determine $(q^1\ldots q^N,\dot q^1\ldots \dot q^N)$, one must 
first choose a particular parametrization; i.e. a particular trajectory 
from the equivalence class $[\gamma]$. There are many ways to do this, 
and hence many different sets of initial data 
$(q^1\ldots q^N,\dot q^1\ldots \dot q^N)$ all denoting the same point 
$[\gamma]$ in phase space. 

Having many different labels for the same point in phase space may be 
somewhat confusing. Given two sets of initial data, it may be difficult  
to determine whether they represent the same point in phase space or 
different points\footnote{One would actually have to solve the equations of 
motion to decide.}. This problem also arises in general relativity, 
where it is difficult to tell if two different sets of initial data 
on a hypersurface $\Sigma$ will generate space-times with the same 
4-geometry. 

The confusion described above arises when one uses initial 
data to parametrize phase space. It clears when one recalls
that these initial data are merely labels, and focuses instead on the 
geometric picture of the phase space discussed above.  

This approach is particularly helpful in the case of general relativity. 
Here $\phase$ is the space of solutions to Einstein's field equations, 
and $\red$ is obtained by factoring out the gauge group; i.e. 
by identifying solutions that can be mapped into each other by 
space-time diffeomorphisms. 

Each point in $\red$ therefore represents a space-time with a Lorentz 
metric satisfying Einstein's equations; for simplicity, we assume here 
that there are no matter fields, although these could easily be incorporated. 
(By ``space-time'', we mean here an equivalence class of isometric 
inextendible connected Hausdorff $C^\infty$ 4-manifolds with $C^2$ Lorentz 
metrics \cite{HawkingEllis}. Each space-time thus represents an orbit of 
the group of diffeomorphisms.) 
One could refine this definition by imposing additional conditions on 
the class of admissible space-times, such as strong causality 
or asymptotic flatness at spatial infinity, but such possibilities 
will not be considered here. 

This definition of the phase space of general relativity is essentially 
that proposed by Witten and Crnkovi\'c \cite{Witten,Crnkovic}. 
Its primary advantage is that it is fully space-time covariant, 
and is formulated without reference to any particular coordinate 
system or preferred time coordinate. 

Of course, in order to describe the physical properties of a given space-time 
one requires a system for describing a particular solution by a set of numerical 
labels. As remarked earlier, a convenient and popular way to do this is by 
specifying initial data on some spacelike initial hypersurface $\Sigma$. 
For example one might specify the induced 3-metric and the second fundamental 
form on $\Sigma$, as in the ADM formulation of canonical general relativity
\cite{ADM}.   
Alternatively, one might follow the approach of Ashtekar and specify the 
self-dual part of the $SO(3)$ connection, along with the spatial triad density 
(with weight one) \cite{Ashtekar}. Either choice of initial data is generally 
sufficient to single out a particular solution of Einstein's equations 
(up to possible isometries). 

However, one is not obliged to use 
initial data to identify points in the phase space of general relativity; 
indeed, solutions of Einstein's equations can be described in entirely 
different ways. For example, Landi and Rovelli propose using the 
eigenvalues of the Dirac operator to identify particular solutions of 
Einstein's equations, and thus to label points in the phase space of 
\cite{RovelliLandi}. Undoubtedly, many other 
approaches also exist.  

\sn{Observables in General Relativity} 
\def\point{{\gamma}}

The central task in canonical quantization is to find operator 
representations for observables. Before attempting to  
quantize a theory, it is therefore important to know which quantities 
qualify as observables. An appropriate definition was given by Dirac
\cite{Dirac}, and is presented below in the terminology of the 
previous section. 

According to Dirac's definition, a function $F:\ext\mapsto\real$ 
on the extended phase space is an observable if its Poisson bracket 
with each first-class constraint $\Phi_a$ vanishes on the constraint 
surface $\phase \subset\ext$:
\begin{equation}
\left.\{\Phi_a,F\}\right|_\phase = 0.    \label{eq:DiracDef} 
\end{equation}
This requirement ensures that $F$ is gauge-invariant, since these 
constraints are the generators of the gauge transformations. Indeed 
we recall that $\{\Phi_a,F\} \equiv -i_{V_a} dF$ where $V_a$ is the 
vector field associated with $\Phi_a$, so if (\ref{eq:DiracDef}) 
holds then 
\begin{equation} 
\left. i_{V_a} dF \right|_\phase =0 
\end{equation}
and hence $F$ is unaffected by gauge transformations. 

An observable $F$ therefore takes the same value at all points in a 
given orbit of the gauge group $\group$, and may be regarded as a 
function on the quotient space $\phase/\group$; that is, on the 
reduced phase space $\red$. The converse is also true; any well-defined 
function on $\red$ can be viewed as a gauge-invariant function on $\phase$ 
and hence as a Dirac observable. To put it another way, Dirac's criterion 
is trivially satisfied by any function $F:\red\mapsto\real$ as there are no 
constraints on $\red$. 

If we adopt the definitions of $\phase$ and $\red$ proposed in the previous 
section, then an observable $F$ will assign a single real value to an entire 
solution $\gamma$ of the equations of motion (and the same value to all other 
solutions obtained from it by gauge transformations). This value does not 
evolve in time, as the state of the system is always represented by the same 
point $[\gamma]$ in the phase space.

However, another kind of time evolution can be observed if the observable $F$ 
is replaced by a family of observables $\{F_t | t\in \real\}$, each of which 
assigns to the entire classical solution $\gamma$ some gauge-independent 
quantity associated with the instantaneous properties of this solution at 
time $t$. For example, if $\gamma:\real\mapsto \real^N$ represents some 
classical particle trajectory, then $F_t$ might assign to this entire 
trajectory (and to gauge equivalent trajectories) a single value determined 
by the instantaneous position and velocity of the particle at time $t$; 
\begin{equation} 
F_t([\gamma]) = f(\gamma(t),\dot\gamma(t)). 
\end{equation}  
Thus, for a fixed choice of $t$, $F_t([\gamma])$ is a single number associated 
with the entire trajectory $\gamma$. On the other hand, if the value of $t$ is 
changed then $F_t$ denotes a different observable and produces a different 
value when acting on $\gamma$. Thus it is the observable itself that 
changes with time, not merely its value. 

This is precisely analogous to the Heisenberg picture of quantum mechanics, 
in which operators evolve rather than state vectors. On the other hand, the 
conventional approach corresponds to the Schr\"odinger picture, in which 
it is the states rather than the operators which evolve. 

This approach is useful when considering general relativity, for 
which the gauge transformations are diffeomorphisms. In this case, an 
observable is a function on $\red$ that assigns to each space-time a 
value that is unaffected by diffeomorphisms; in other words, 
a geometric invariant of the space-time manifold. 

In pure general relativity, in the absence of any preferred coordinate 
system, it is difficult to identify a particular space-time point at which 
local data is to collected. For this reason, geometric invariants of 
space-time must generally be defined in a global manner without reference 
to any special space-time points. 
These globally defined quantities may be expressible as integrals of local 
invariants such as $R$, $R^2$, $R^{\mu\nu}R_{\mu\nu}$ or 
$C^{\mu\nu\rho\sigma}C_{\mu\nu\rho\sigma}$ (where $R, R_{\mu\nu}$ and
$C_{\mu\nu\rho\sigma}$ denote the Ricci scalar and the Ricci and Weyl
tensors). There are also a variety of globally defined geometric 
invariants (such as the eigenvalues of the Dirac or Klein-Gordon operators 
\cite{RovelliLandi}) that cannot be expressed simply as integrals of local 
quantities.

While all of these are genuine Dirac observables, none can be evaluated 
without full knowledge of the future and past of the Universe. In 
particular, their values cannot be ascertained by a localized observer 
who only has access to data from his causal past (unless the observer 
inhabits a deterministic space-time \cite{BudicSachs}). 

Real observations are made locally, and the things we can measure 
are   {\it not} globally defined geometric invariants of the 
space-time manifold as described above. 
Observables of this type are therefore inappropriate quantities to 
consider in a theory which purports to relate to physical observation. 
What is clearly needed is a set of observables whose values are 
determined by local properties associated with particular points or 
regions in space-time. The difficulty is that any such local observables 
must be unaffected by arbitrary diffeomorphisms that map points to 
different points. 

The only way around this problem would be to evaluate local invariants 
at space-time points that are identified in a diffeomorphism-invariant 
manner; for example, as the points at which certain local invariants
take specified values. However, this approach does not appear to work 
in the case of pure gravity, since at each point there are only 4 
algebraically independent local invariants (all obtained from the Weyl tensor 
$C_{\mu\nu\rho\sigma}$ \cite{Weinberg}). Hence, even if a point could 
be identified as the unique location where these 4 quantities took 
specified values, there would be no remaining independent local 
invariants to measure there.  

If matter is present then there are more possibilities. 
For example, Rovelli proposes labelling space-time points 
with reference to a ``material reference system'' 
provided by a space-filling cloud of particles carrying clocks 
\cite{RovelliMRS}. 
He shows how this permits the definition of observables that are 
local but also invariant under space-time diffeomorphisms. However, 
he emphasizes that the material nature of the reference particles 
is a vital ingredient; not only will their trajectories in space-time 
be determined by appropriate equations of motion, but they will also 
have a non-vanishing effect on the geometry of space-time to which 
they must be coupled via Einstein's equations.

While this approach provides a large set of local Dirac observables for 
general relativity coupled to matter, it has some shortcomings. In the 
first place, it requires the existence of a collection of 
particles\footnote{These particles must have non-zero rest mass, as the 
trajectories of massless particles cannot be parametrized by proper time.} 
with clocks and consequently sheds 
no light on the problem of identifying observables in a theory without matter. 

More importantly, observation is ultimately a local process and 
data obtained from the various particles does not constitute an 
observation until it has been collected by a single localized 
observer. Attention must therefore be focussed on this observer 
and the set of observables to which he has access; data from the 
region outside his causal past cannot form part of any observation.

\sn{Space-Times with Localized Observers}
\def\ob{{\cal O}}
\def\J{{\cal J}} 
\def\N{{\cal N}} 

In practice, all physical theories must deal with observations that can 
be collected by a single localized observer\footnote{This point is of 
critical importance in quantum mechanics, where the local nature of 
observation must be recognized if one is to avoid the causal paradoxes 
associated with the instantaneous collapse of the wave function.}. 
Indeed, the scientific method requires that a theory should be subjected 
to experimental tests that culminate in the collection and interpretation 
of data by a scientist or a localized team of scientists. 

Although Rovelli's approach allows us to define a variety of 
local observables, many of these will be inaccessible 
to a localized observer. Moreover, such an observer will be unable 
to determine the value of any globally defined observables. 
This argument suggests that the observer should be incorporated 
into the definition of an observable. We therefore start with some 
definitions. 

According to Hawking and Ellis \cite{HawkingEllis}, a space-time 
can be represented by a $C^2$-inextendible pair $({\cal M},{\bf g})$ 
where ${\cal M}$ is a connected 4-dimensional Hausdorff $C^\infty$ 
manifold and $\bf g$ is a $C^2$ Lorentz metric on $\cal M$. 
We now modify this definition by requiring each space-time to contain 
a privileged point corresponding to the observer's ``here and now''; 
a space-time will then be represented by a triple $({\cal M},\ob,{\bf g})$ 
where $\ob \in \cal M$ represents this special point. The same 
space-time can also be represented by any other triple related to 
$({\cal M},\ob,{\bf g})$ by an isometry, but it is generally 
convenient to work with a single representative triple. 

For the sake of clarity, we restrict the following analysis to 
pure gravity. However, it is a simple matter to accommodate matter 
fields if desired.  
 
For any $({\cal M},\ob,{\bf g})$ the {\it observer's causal past} is 
defined as the set $\J^-(\ob,{\cal M})$ of all points in $\cal M$ that 
can be connected to $\ob$ by 
future-directed non-spacelike curves in $\cal M$. 
From the perspective of a localized observer performing an experiment 
that culminates at $\ob$, the set $\J^-({\ob},{\cal M})$ represents 
the visible part of space-time. 
This motivates us to focus on developing a classical theory 
that deals exclusively with the parts of space-time that are visible 
from $\ob$  -- without necessarily regarding them as part of some 
larger hypothetical\footnote{The 
work ``hypothetical'' is appropriate because nothing can be known 
about the region outside the observer's causal past.} space-time. 

Two space-times represented by triples $({\cal M},\ob,{\bf g})$ and 
$({\cal M}',\ob',{\bf g}')$ will be said to be {\it indistinguishable} 
if the interior of $\J^-(\ob,{\cal M})$ is isometric to the interior 
of $\J^-(\ob',{\cal M}')$.
This is clearly an equivalence relation, which we denote $\sim$.  
An equivalence class of indistinguishable space-times will be referred 
to as a {\it visible space-time}, and characterizes the geometry of the 
observer's causal past; it is what remains of a space-time when we  
discard what cannot be observed. If this equivalence class includes 
at least one space-time on which the metric satisfies Einstein's 
equations, then it will be referred to as a {\it classical} 
visible space-time. 

In case the above definition seems rather formal, note that a 
visible space-time can also be represented by a pair $(\J,{\bf g})$ 
where  
\begin{itemize} 
\item $\J$ is a connected four-dimensional manifold-with-boundary, 
and is smooth everywhere except at a singular point $\ob\in\J$;
\item $\ob$ has a neighbourhood $\N\subset\J$ on which is defined a 
homeomorphism $\phi:\N\to \{(x^1,x^2,x^3,x^4)\in\real^4 \, : \, 
(x^1)^2+(x^2)^2+(x^3)^2 \leq (x^4)^2, \, x^4 \leq 0\}$ 
which is smooth on $\N\backslash\ob$ and also regarded as 
smooth\footnote{The homeomorphism $\phi$ defines a local coordinate 
system on $\N$, and extends the differential 
structure on $\J\backslash\{\ob\}$ to cover the point $\ob$. 
Thus, a function $f:\N\to\real$ is said to be $C^k$ at $\ob$ if 
$f\circ\phi^{-1}$ is $C^k$ at $\phi(\ob)=(0,0,0,0)$.}  
at $\ob$, with $\phi(\ob)=(0,0,0,0)$;
\item $\bf g$ is a $C^2$ Lorentz metric on ${\J}$;
\item every point in ${\J}\backslash\{\ob\}$ can be connected to $\ob$ 
by a non-spacelike curve in ${\J}$;
\item $(\J,{\bf g})$ is $C^2$-inextendible, in the sense that 
it is not isometric to a proper subset of another pair $(\J',{\bf g'})$ 
with the above properties.  
\end{itemize} 

Note that the causal past of any interior point of space-time 
has precisely these properties. If the metric $\bf g$ satisfies Einstein's 
equations everywhere on $\J$, the visible space-time is classical.  

In section 2 we showed how the reduced phase space $\red$ is obtained 
by discarding the unphysical gauge degrees of freedom from 
a larger phase space. In particular, for general relativity we saw 
that elements of $\red$ are space-times with metrics satisfying Einstein's 
equations. However these space-times still contain unphysical degrees of 
freedom; those which the observer cannot measure because they are 
associated with points outside his causal past. To remove these 
unphysical degrees of freedom, we therefore define 
the reduced phase space $\bar\red$ of our 
observer-based theory as the quotient space $\bar\red=\red/\sim$
obtained from $\red$ by identifying indistinguishable space-times.
This quotient space inherits a symplectic structure from $\red$.  
(The inherited symplectic form will be non-degenerate because 
there is no gauge group acting on $\bar\red$.) The elements of 
$\bar\red$ are then the classical visible space-times. 

Unlike $\red$, the new reduced phase space $\bar\red$ is partially 
ordered; given two elements of $\bar\red$ represented by triples 
$({\cal M},\ob,{\bf g})$ and $({\cal M}',\ob',{\bf g}')$ respectively, 
we say that the first {\it contains} the second if the interior of 
$\J^-(\ob',{\cal M}')$ is isometric to an open subset of 
$\J^-(\ob,{\cal M})$. It is easy to verify that the relation of 
containment is reflexive, antisymmetric and transitive, and therefore 
a partial ordering.  
 
The physical meaning of this ordering relation is straightforward; if one 
visible 
space-time contains another, it means that the causal past of the first 
observer can be regarded as containing the causal past of the first, and 
so the first observer may be regarded as being in the casual future of 
the second.  
It follows that any future-directed causal curve in space-time correpsonds 
to a totally ordered subset of the reduced phase space $\bar\red$; 
conversely, any totally ordered subset of $\bar\red$ corresponds to
a set of points along a future-directed causal curve. 
 
We now investigate what observables the theory admits. These are defined 
as functions on the reduced phase space $\bar\red$. Thus, an 
observable is a rule assigning a real quantity 
to each classical visible space-time; i.e. a geometric invariant of 
the observer's causal past.

One such observable is the 4-volume of the observer's causal past (if this 
happens to be finite). In fact, the value of this observable is strictly 
increasing along the observer's world-line, and is therefore 
naturally regarded as a time parameter. Its status as a bona fide 
observable suggests that it may be used in constructing a time-dependent 
version of the quantum theory. (Note that this quantity cannot be 
defined in the conventional formulation of general relativity, in which 
there is no privileged point representing the observer's here and now.) 

It also proves quite easy to find local observables in this theory, 
as one can readily locate space-time points with reference to the observer's 
here-and-now $\ob$. For example, by constructing a Riemann normal coordinate 
system about $\ob$, it is possible to attach a label to every point in a 
neighbourhood of $\ob$ with a mimimum of ambiguity; the only arbitrariness 
in this procedure is that associated with the $SO(1,3)$ freedom available 
in choosing the directions of the coordinate axes. One can then take as 
observables the values of any local invariants at specified points in 
this neighbourhood. 

The conclusion of this analysis is that any measurements of local 
invariants collected from points nearby the observer in his causal past
qualify as observables. This coincides very closely with what most 
physicists mean by observables.

\sn{Boundary Data and Quantization}

In order to put this theory into a more conventional form, we 
consider what boundary data must be specified in order to 
identify particular elements of the reduced phase space $\bar\red$; 
i.e. a classical visible spacetime. 

A classical visible space-time corresponds to a possible geometry 
of the observer's causal past $\J^-(\ob)$.  
In fact the causal structure of general relativity ensures that 
this geometry can be reconstructed using Einstein's equations from final 
data on the observer's past light cone $C^-$, defined here as 
the null surface generated by past-directed null geodesics 
through $\ob$. (We are only considering pure general relativity, and 
therefore don't need to worry about the possible formation of caustics.) 
Indeed, Dautcourt 
has shown that this can be achieved with knowledge of just two 
real function of the metric on $C^-$ \cite{Dautcourt}. 
Suppose  the metric is written in the form 
\begin{equation}
ds^2=m^2\,du^2 + 2h\, du\, dv \, + 2k_A w^A du \, +g_{AB} dw^A dw^B
\end{equation}
where $u$ and $v$ are null coordinates, and $w^A$ ($A=2,3$) 
are constant along the null generators of $C^-$, and 
$m^2 = k_Ak_B g^{AB}$. Then it is 
sufficient to specify two of the three independent components 
of $g_{AB}$ everywhere on $C^-$; for example, one might 
specify the conformal part of $g_{AB}$ (i.e. $g_{AB}$ up to 
a conformal factor). 

This also happens to be just enough data to specify the intrinsic 
geometry of $C^-$. A metric on a 3-surface generally has 6 independent 
components at each point; however, $C^-$ is null so the metric is 
degenerate and only has 5. Of these, 3 can be removed by an appropriate 
choice of coordinates, leaving just 2 degrees of freedom at each point 
in $C^-$ corresponding, for example, to the components of 
$(\det[g_{AB}])^{-1/2} g_{AB}$. 

It therefore appears that the entire 4-geometry of the observer's 
causal past can be reconstructed from a knowledge of the 
intrinsic geometry of his past light cone. It is therefore 
natural to use this boundary data to label elements of 
the reduced phase space $\bar\red$. 

It should also be possible to represent the symplectic form on 
phase space as an antisymmetric bilinear function 
\begin{equation}
\sympform(\tilde \delta g_{AB},\tilde \delta g'_{CD}) 
 = 
\int_{C^-}  \tilde\delta g_{AB} \sympform^{ABCD} \tilde\delta g'_{CD}
\end{equation}
where the 3-form $\sympform^{ABCD}=-\sympform^{CDAB}$ scales like 
$(\det[g_{AB}])^{-1}$ under conformal transformation of $g_{AB}$, and  
\begin{equation} 
\tilde\delta g_{AB} \equiv \delta g_{AB} - \frac{1}{2} g_{AB} g^{CD} 
\delta g_{CD}. 
 \end{equation} 
In principle it should be possible to derive an expression for 
$\sympform^{ABCD}$ using the results of \cite{Crnkovic}, 
but this will not be attempted here. 

We conclude with a brief discussion of the quantum theory. Prior to 
quantization, one must identify a set of generalized coordinates $q^\alpha$ 
and momenta $p_\alpha$ such that the symplectic form can be written as 
\begin{equation}
\sympform = dp_\alpha \wedge dq^\alpha. 
\end{equation}  
The generalized coordinates $q^\alpha$ will contain just half the 
degrees of freedom needed to parametrize the reduced phase space, 
and thus represent a single degree of freedom of $g_{AB}$ at each 
point in $C^-$. The momenta $p_\alpha$ contain the remaining degree 
of freedom at each point in $C^-$.

As there are no first-class constraints in the reduced phase space, 
the wave function may be taken as an arbitrary complex function 
$\Psi(q)$ of the generalized coordinates $q^\alpha$, with the 
generalized momenta $p_\alpha$ represented as differential operators. 
In order to select a particular wave-function, it would be 
necessary to augment the canonical theory with an appropriate 
set of boundary conditions arising from non-dynamical considerations. 

As remarked in the previous section, one of the observables 
is strictly increasing along any classical world line (i.e. 
any totally ordered sequence in the reduced phase space) and 
is therefore naturally regarded as a time parameter. In the 
quantum theory, this observable will be represented by a 
Hermitian operator on the space of wave functions. The 
eigenstates of this operator will represent quantum states 
in which the observed Universe has a definite age, and so 
projecting the wave function into these eigenstates will 
result in a time-dependent version of the quantum theory.  

\sn{Summary and Discussion} 
By explicitly incorporating a localized observer into the canonical 
analsysis of general relativity, and eliminating the unobservable 
degrees of freedom, we have obtained a modified reduced phase space 
equipped with a natural partial ordering that encapsulates the notion 
of causality. The advantage of this approach is that the set of 
observables now includes the results of local measurements, as well as 
a natural time parameter. 

The analysis suggests that elements of the reduced phase space 
can be identified by specifying the intrinsic geometry of the 
observer's past light cone $C^-$. Once suitable gauge conditions have 
been imposed, this means specifying two real quantities at each 
point of $C^-$. It is expected that these can be chosen so that one 
may be thought of as a generalised coordinate, and the other as a 
generalized momentum; in the quantum theory, the latter will be 
represented by a functional derivative.  

Because the reduced phase space contains no gauge degrees of 
freedom, there are no constraints on the form of the wave function 
(except possibly for boundary conditions that might arise from
non-dynamical considerations). 

The motivation for this approach is the recognition that observation 
is a local phenomenon, and that an observable should therefore 
correspond to physical data that can collected at a definite time 
and place. The price that must be paid is a partial loss of determinism; 
the theory does not permit predictions about the future 
although it does allow deductions about the past. However, unless one 
can justify the imposition of suitable boundary conditions outside the 
observer's causal past, loss of determinism is inevitable and realistic. 

The same approach can of course be applied to theories other than 
general relativity, but this leads to only very minor modifications.
Even in the conventional approach these theories admit local observables 
and a natural time parameter, so no advantage is gained by using the 
approach described here. Moreover, non-gravitational fields 
can effectively be shielded and so it is generally reasonable to 
assume that they will satisfy boundary conditions on surfaces 
outside the observer's causal past. In such cases the observer 
effectively has information about the behaviour of the fields 
outside the region he can observe directly, and so the 
dynamics may be regarded as deterministic. It is only in the 
case of gravity that one is forced to address the indeterminacy 
of a theory dealing with the observations made by a localized 
observer.

\end{document}